# Alternating optimization for G×E modelling with weighted genetic and environmental scores: examples from the MAVAN study

**Alexia Jolicoeur-Martineau[1], Ashley Wazana[2 3 8], Eszter Szekely[4], Meir Steiner[5], Alison S. Fleming[6], James L. Kennedy[6 7], Michael J. Meaney[8 9 10], Celia M.T. Greenwood[9 11 12 13 14] and the MAVAN team.**

[1]Jewish General Hospital, Montreal, Qc, Canada; [2]Centre for Child Development and Mental Health, Jewish General , Montreal, Qc, Canada; [3]McGill University, Montreal, Qc, Canada; [4]Department of Psychiatry, Faculty of Medicine, McGill University, Montreal, Qc, Canada; [5]McMaster University and St-Joseph's Healthcare, Hamilton, On, Canada; [6]University of Toronto, Toronto, On, Canada; [7]Center for Addiction and Mental Health, Toronto, On, Canada; [8]Douglas Mental Health University Institute, Montreal, Qc, Canada; [9]Ludmer Centre for Neuroinformatics and Mental Health, Montreal, Qc, Canada; [10]Sackler Program for Epigenetics & Psychobiology, McGill University, Montreal, Qc, Canada; [11]Lady Davis Institute for Medical Research, Jewish General Hospital, Montreal, Qc, Canada; [12]Department of Epidemiology, Biostatistics and Occupational Health, McGill University, Montreal, QC, Canada; [13]Department of Oncology, McGill University, Montreal, Qc, Canada. [14]Department of Human Genetics, McGill University, Montreal, Qc, Canada.

## Abstract

Motivated by the goal of expanding currently existing genotype × environment interaction (G×E) models to simultaneously include multiple genetic variants and environmental exposures in a parsimonious way, we developed a novel method to estimate the parameters in a G×E model, where G is a weighted sum of genetic variants (genetic score) and E is a weighted sum of environments (environmental score). The approach uses alternating optimization to estimate the parameters of the G×E model. This is an iterative process where the genetic score weights, the environmental score weights, and the main model parameters are estimated in turn assuming the other parameters to be constant. This technique can be used to construct relatively complex interaction models that are constrained to a particular structure, and hence contain fewer parameters.

We present the model as a two-way interaction longitudinal mixed model, for which ordinary linear regression is a special case, but it can easily be extended to be compatible with k-way interaction models and generalized linear mixed models. The model is implemented in R (*LEGIT* package) and using SAS macros (*LEGIT_SAS*). Here we present examples from the Maternal Adversity, Vulnerability, and Neurodevelopment (MAVAN) study where we improve significantly upon already existing models using alternating optimization. Furthermore, through simulations, we demonstrate the power and validity of this approach even with small sample sizes.

# Introduction

## Genotype × environment interactions

In the past few decades, genotype × environment interaction (G×E) models have been widely used in Epidemiology, Medicine, and Psychology (Meaney, 2010, Caspi and Moffitt, 2006, Belsky et al., 2009). Instead of partitioning variance into separable and independent genetic and environmental contributions to phenotypic differences, G×E models assume the interdependence of genetic and environmental influences on a given trait (Rutter, 2007). Consequently, the genes and the environment may not necessarily have independent effects on the phenotype, it is rather their interaction that matters most. The G×E model is consistent with findings in Molecular Biology that the activation of gene expression is contingent upon transcriptional signals that derive from the internal and the external environment (Meaney, 2010).

## The need for multiple genetic and environmental factors

The simplest G×E models are generally represented as ordinary linear models including a single genetic variant and a single environmental exposure, as expressed by the following formula:

$$\boldsymbol{y} = \beta_0 + \beta_e \boldsymbol{e} + \beta_g \boldsymbol{g} + \beta_{ge} \boldsymbol{g}\boldsymbol{e} + \boldsymbol{\varepsilon} \,,$$

where $\boldsymbol{y}$ is the outcome, $\boldsymbol{e}$ is the environment, $\boldsymbol{g}$ is the genetic variant, $(\beta_0, \beta_g, \beta_e, \beta_{ge})$ are the parameters to be estimated and $\boldsymbol{\varepsilon}$ is the error term. Examining the effect of only a single genetic variant and an environmental exposure at a time significantly limits the potential explanatory value of the model. Furthermore, if one intends to study the effect of multiple genes and/or environmental factors, often separate models are built which do not take into account the various interactions between all the genes and environments considered. These individual-variable models often have very small effect sizes and low replication rates (Lee et al., 2012, Risch et al., 2009).

In rare cases, disease pathology can be accurately predicted using a single genetic variant and an environmental variable. An example of this would be Phenylketonuria (Al Hafid and Christodoulou, 2015), a disorder that increases the levels of phenylalanine in the blood. Homozygous carriers of the hepatic enzyme phenylalanine hydroxylase (PAH) gene have the disease, but as long as they follow a phenylalanine-free diet from birth, they will remain symptom-free (Al Hafid and Christodoulou, 2015). The phenotype only manifests itself in the presence of a medium-high phenylalanine diet. With complex traits, like most psychological and mental health outcomes, a single genetic variant and environment are rarely sufficient to explain a significant proportion of the phenotypic variance.

## The genetic and environmental scores

Although there exist many methods to incorporate multiple genetic variants and environments into a single model, here we focus on the use of genetic and environmental scores. These scores are assumed to be linear combinations of individual (genetic or environmental) variables and to represent latent variables capturing the overall genetic and environmental contributions to the phenotype. We consider the situation where a model includes interactions between these scores, thereby making some specific but implicit assumptions about the form of interactions between the individual genetic variants and environment. This restrains the model to a very small subset of the full potential space of G×E, but in doing so, assigns a biologically plausible structure.

Assuming we have a selection of genetic variants $\boldsymbol{g_1}, \dots, \boldsymbol{g_k}$ (from a few SNPs to a full genome) and environments of interest $\boldsymbol{e_1}, \dots, \boldsymbol{e_s}$, we define genetic and environmental scores as:

$$\boldsymbol{g} = \sum_{j=1}^{k} p_j \boldsymbol{g_j}\ , \qquad (1)$$

$$\boldsymbol{e} = \sum_{l=1}^{s} q_l \boldsymbol{e_l}\ ,$$

where $\boldsymbol{p} = (p_1, \dots, p_k)$ is a vector of unknown parameters creating a score from a linear combination of the genetic variables and $\boldsymbol{q} = (q_1, \dots, q_s)$ is a vector of unknown parameters that similarly creates a score from the environmental variables.

Given the categorical nature of genetic variants (e.g. for two alleles, A & a, the genotypes are AA, Aa, aa), there are multiple ways to code them as variables for use in a statistical model. After choosing one allele of interest (usually the least common allele, also called the minor allele), a common coding choice assumes that each copy of the allele of interest has an additive effect (e.g. 0=aa, 1=aA, 2=AA). Another common coding choice is to create 2 binary variables (e.g. 0=aa, 1=aA, 1=AA or 0=aa, 0=aA, 1=AA) to capture the differences between the 3 genotypes. Environmental exposures are commonly assessed using questionnaires and coded as ordinal (0, 1, 2 …) or continuous variables. Our alternating optimization approach makes no distinction on the choice of variable coding used; in fact, any potentially interacting variables, not necessarily genetic or environmental, could be used.

Note that our focus is on how to best construct scores inside G×E models assuming we have already selected the specific genetic variants and environments to include. The topic of variable selection is not the subject of the paper, although some information on variable selection with this approach is available in Appendix B.

### How genetic scores are traditionally constructed

A rather varied terminology has been used in the literature to refer to genetic scores, such as “multilocus genetic profile” (Green et al., 2016, Nikolova et al., 2011), “allelic score” (Burgess and Thompson, 2013, Spycher et al., 2012), “SNP score” (Vrieze et al., 2012), “genotype score” (Meigs et al., 2008), “genetic prediction score” (Zhao et al., 2014) and most commonly “polygenic risk score” (Abraham et al., 2013, Cho et al., 2010, de Vlaming and Groenen, 2015, Mak et al., 2016).

Genetic variants can be included in the genetic score either based on a hypothesis-free approach or a hypothesis-driven approach (Belsky and Israel, 2014). The first one is most commonly based on a genome-wide search of associated signals (Sullivan, 2010), possibly using a p-value cutoff as indication for association (Stergiakouli et al., 2016). The second approach is based on the theoretical biological understanding of the phenotype studied.

Although rarely used nowadays, one approach to constructing genetic scores is to assign an equal weight for each locus (Green et al., 2016, Nikolova et al., 2011). This can be seen as an important limitation, as it is biologically rather unlikely that every additional risk allele at every locus considered makes an equal contribution at the molecular level. On the other hand, some studies have found little to no improvement when using unequal weights in their analysis (Machiela et al., 2011) although this might have been due to using non-optimal weights. Importantly, the direction of the weights can also change depending on the specific loci included and any unaccounted gene × gene interactions (G×G) may also influence the direction and magnitude of a gene's effect.

The most common way of assigning weights to genetic variants is to use effect sizes observed for those variants in independent genome-wide association studies (GWAS) (Belsky and Israel, 2014). Alternatively, regularization (penalization) techniques like lasso (Tibshirani, 1996), ridge (Hoerl and Kennard, 1970) or elastic net (Zou and Hastie, 2005) are used for estimating the weights of the genetic variants in an independent, discovery sample (Abraham et al., 2013, Cho et al., 2010, de Vlaming and Groenen, 2015, Mak et al., 2016). Although estimating genetic weights a priori in independent samples provides less biased scores, the vast majority of available studies report only main effects. If the association of some genetic variants depends on the environment, these discovery sample estimates will represent the average effect across different levels of the environment.

It is also possible to construct genetic scores when the phenotype studied is different from that in the discovery sample. For example, discovery samples of specific medical disorders are often used to study associated symptoms (Musliner et al., 2015), continuous traits (Derks et al., 2012, Martin et al., 2014) or age of onset (Chibnik et al., 2011, Nalls et al., 2015). Intermediate phenotypes are sometimes also used to generate risk scores for clinical phenotypes (Fontaine-Bisson et al., 2010, Horne et al., 2005). The weights used for these types of analysis are likely to be suboptimal for the new phenotypes.

For a complete review of the different approaches to construct genetic scores and their use in G×E models, please see Belsky and Israel (2014).

## How environmental scores are traditionally constructed

There have been prior attempts to combine multiple environmental exposures into a cumulative environmental risk score with limited complexity. These methods have generally dichotomized the environments as "good" or "bad" and combined them into an ordinal composite scale representing a cumulative environmental effect; for example, Sameroff's environmental risk scale (Sameroff, 1998) and Adverse Childhood Experiences (ACE) score (Felitti et al., 1998). A more recent approach to create environmental scores is based on environment-wise association studies (EWAS) (Park et al., 2014), however, this is still a very new area of research. The traditional practice is to simply run separate models for each environment considered.

Here, we present a novel approach referred to as alternating optimization, which estimates the weights of the genetic and environmental score in equation (1) at the same time as the interactions between the scores.

## Methods

### Alternating optimization

The principal idea behind alternating optimization is to construct a complex model in parts, rather than constructing one big model and estimating all possible main effects or interactions at the same time. In one part of the model, we estimate the weights of the genetic score while holding the other parameters constant. In a second part, we estimate the weights of the environmental score while holding the other parameters constant. In the final part of the model, we estimate the parameters for the main effects and interaction effects of the genetic and environmental scores while holding these scores constant. Finding an optimal solution for the parameters of the full model can be challenging (non-convex, non-linear, high dimensionality) whereas finding the locally optimal solution for each of the three parts of the models is relatively straightforward.

We present the basic idea in the context of a two-way interaction model with normal errors. However, in the Discussion, we provide information on how to adapt this technique to estimate main effects, three-way effects, k-way effects models or generalized linear mixed models (GLMM) with non-identity link functions.

Assuming a two-way interaction between the genetic score $\boldsymbol{g}$ and the environmental score $\boldsymbol{e}$, the model can be defined as:

$$\boldsymbol{y} = \beta_0 + \beta_e\boldsymbol{e} + \beta_g\boldsymbol{g} + \beta_{eg}\boldsymbol{eg} + \boldsymbol{X}_{covs}\boldsymbol{\beta}_{covs} + \boldsymbol{\varepsilon}\,, \tag{2}$$

where $\boldsymbol{y}$ is a vector representing the $n$ observed outcomes, $\beta_0, \beta_e, \beta_g, \beta_{eg}$ are scalars of the unknown parameters for the G×E, $\boldsymbol{X}_{covs}$ is a design matrix for additional covariates, $\boldsymbol{\beta}_{covs}$ a vector of unknown parameters for the covariates and $\boldsymbol{\varepsilon}$ is the error term.

Within model (2), there are infinitely many possibilities for $\boldsymbol{p} = (p_1, \ldots, p_k)$ or $\boldsymbol{q} = (q_1, \ldots, q_k)$ that lead to the same fit and p-values. This can be best illustrated by the fact that $\beta_g, \beta_{eg}$ with $c\boldsymbol{p}$, where $c$ is a constant, leads to the exact same model as $c\beta_g, c\beta_{eg}$ with $\boldsymbol{p}$:

$$\begin{aligned}\boldsymbol{y} &= \beta_0 + \beta_e\boldsymbol{e} + \beta_g\sum_{j=1}^{k}(cp_j)\boldsymbol{g_j} + \beta_{eg}\boldsymbol{e}\sum_{j=1}^{k}(cp_j)\boldsymbol{g_j} \\ &= \beta_0 + \beta_e\boldsymbol{e} + c\beta_g\sum_{j=1}^{k}p_j\boldsymbol{g_j} + c\beta_{eg}\boldsymbol{e}\sum_{j=1}^{k}p_j\boldsymbol{g_j} \\ &= \beta_0 + \beta_e\boldsymbol{e} + (c\beta_g)\boldsymbol{g} + (c\beta_{eg})\boldsymbol{eg}\end{aligned}$$

Therefore, to prevent infinite possibilities for $\boldsymbol{p}$ and $\boldsymbol{q}$, we add the following restrictions: $\|\boldsymbol{p}\|_1 = \sum_{j=1}^{k}|p_j| = 1$ and $\|\boldsymbol{q}\|_1 = \sum_{l=1}^{s}|q_l| = 1$ to the genetic and environmental scores in (1).

These restrictions also provide a very helpful interpretation; the absolute value of $p_j$ represents the relative contribution of the $j^{th}$ genetic variant to the genetic score and the sign of $p_j$ represents the direction of the $j^{th}$ genetic variant's contribution to the genetic score $\boldsymbol{g}$. A similar set of restrictions is applied to the environmental score parameters. Although with these restrictions it is theoretically possible to find the true absolute values of the model parameters, the true signs of the parameters are still unknown since the restrictions do not force any sign. This is discussed in more detail in Appendix B.

**Estimation**

Because directly estimating the parameters $(\boldsymbol{\beta}, \boldsymbol{p}, \boldsymbol{q})$ still remains a challenge, we apply alternating optimization to reduce the complexity of the problem. More specifically, instead of estimating $(\boldsymbol{\beta}, \boldsymbol{p}, \boldsymbol{q})$ simultaneously, we first estimate $\boldsymbol{\beta}$ while holding $\boldsymbol{p}$ and $\boldsymbol{q}$ constant, then we estimate $\boldsymbol{p}$ while holding $(\boldsymbol{\beta}, \boldsymbol{q})$ constant and finally we estimate $\boldsymbol{q}$ while holding $(\boldsymbol{\beta}, \boldsymbol{p})$ constant. This process is repeated until convergence is obtained. The pseudocode for this algorithm is given below:

**Algorithm 1**: Alternating optimization for estimating the parameters of a two-way G×E model.

1. Set $\widehat{\boldsymbol{p}}^0$ and $\widehat{\boldsymbol{q}}^0$ to reasonable starting points
2. Until convergence, for $i = 1$ to $max_iterations$
   2.1. Estimate $\boldsymbol{\beta}$ assuming $\boldsymbol{p} = \widehat{\boldsymbol{p}}^{i-1}$ and $\boldsymbol{q} = \widehat{\boldsymbol{q}}^{i-1}$
   2.2. Let $\widehat{\boldsymbol{\beta}}^i = \widehat{\boldsymbol{\beta}}$
   2.3. Estimate $\boldsymbol{p}$ assuming $\boldsymbol{\beta} = \widehat{\boldsymbol{\beta}}^i$ and $\boldsymbol{q} = \widehat{\boldsymbol{q}}^{i-1}$
   2.4. Let $\widehat{\boldsymbol{p}}^i = \frac{\widehat{\boldsymbol{p}}}{\|\widehat{\boldsymbol{p}}\|_1}$
   2.5. Estimate $\boldsymbol{q}$ assuming $\boldsymbol{\beta} = \widehat{\boldsymbol{\beta}}^i$ and $\boldsymbol{p} = \widehat{\boldsymbol{p}}^i$
   2.6. Let $\widehat{\boldsymbol{q}}^i = \frac{\widehat{\boldsymbol{q}}}{\|\widehat{\boldsymbol{q}}\|_1}$
   2.7. If $\|\widehat{\boldsymbol{p}}^i - \widehat{\boldsymbol{p}}^{i-1}\| < \delta$ and $\|\widehat{\boldsymbol{q}}^i - \widehat{\boldsymbol{q}}^{i-1}\| < \delta$ then convergence is attained (break loop)
3. Return $(\widehat{\boldsymbol{\beta}}^i, \widehat{\boldsymbol{p}}^i, \widehat{\boldsymbol{q}}^i)$

Step 2.3 and Step 2.5 warrant a little more explanation. To fit the model in Step 2.3, we must reparametrize the model in the following way:

$$\boldsymbol{y}' = \sum_{j=1}^{k} p_j \boldsymbol{r}_1^j + \boldsymbol{\varepsilon}. \tag{3}$$

where $\boldsymbol{y}' = (\boldsymbol{y} - (\beta_0 + \beta_e \boldsymbol{e} + \boldsymbol{X_{covs}}\boldsymbol{\beta_{covs}}))$ and $\boldsymbol{r}_1^j = ((\beta_g + \beta_{eg}\boldsymbol{e})\boldsymbol{g}_j)$. Similarly, to fit the model in Step 2.5, we must reparametrize the model in the following way:

$$\boldsymbol{y}'' = \sum_{l=1}^{s} q_l \boldsymbol{r}_1^{l\,\prime} + \boldsymbol{\varepsilon}. \tag{4}$$

where $\boldsymbol{y}'' = (\boldsymbol{y} - (\beta_0 + \beta_g \boldsymbol{g} + \boldsymbol{X_{covs}}\boldsymbol{\beta_{covs}}))$ and $\boldsymbol{r}_1^{l\,\prime} = ((\beta_e + \beta_{eg}\boldsymbol{g})\boldsymbol{e}_l)$. The derivation of these parameterizations and more details of the algorithm are available in Appendix A.

## Properties

The alternating optimization approach always converges to a local optimum.

Assuming the following notation: $\widehat{\mathbf{\Theta}}^{i} = \left(\widehat{\boldsymbol{\beta}}^{i}, \widehat{\boldsymbol{p}}^{i}, \widehat{\boldsymbol{q}}^{i}\right)$; we denote the in-between steps as $\widehat{\mathbf{\Theta}}^{i+1/3} = \left(\widehat{\boldsymbol{\beta}}^{i+1}, \widehat{\boldsymbol{p}}^{i}, \widehat{\boldsymbol{q}}^{i}\right)$ and $\widehat{\mathbf{\Theta}}^{i+2/3} = \left(\widehat{\boldsymbol{\beta}}^{i+1}, \widehat{\boldsymbol{p}}^{i+1}, \widehat{\boldsymbol{q}}^{i}\right)$. The objective function is denoted as $\mathbf{F}(\widehat{\mathbf{\Theta}})$. We also assume that $\mathbf{F}(\widehat{\mathbf{\Theta}})$ is a function to be maximized (as in the case of ML, REML, ect). Then, we have that the sequence $\left\{\mathbf{F}(\widehat{\mathbf{\Theta}}^{0}), \mathbf{F}(\widehat{\mathbf{\Theta}}^{1/3}), \mathbf{F}(\widehat{\mathbf{\Theta}}^{2/3}), \mathbf{F}(\widehat{\mathbf{\Theta}}^{1}), \mathbf{F}(\widehat{\mathbf{\Theta}}^{1+1/3}), \mathbf{F}(\widehat{\mathbf{\Theta}}^{1+2/3}), \ldots\right\}$ is monotone increasing, i.e: $\mathrm{F}\left(\widehat{\mathbf{\Theta}}^{0}\right) \leq \mathrm{F}\left(\widehat{\mathbf{\Theta}}^{1/3}\right) \leq \mathrm{F}\left(\widehat{\mathbf{\Theta}}^{2/3}\right) \leq \mathrm{F}\left(\widehat{\mathbf{\Theta}}^{1}\right) \leq \mathrm{F}\left(\widehat{\mathbf{\Theta}}^{1+1/3}\right) \leq \mathrm{F}\left(\widehat{\mathbf{\Theta}}^{1+2/3}\right) \leq \cdots \leq$ . Note that if $\mathrm{F}(\widehat{\mathbf{\Theta}})$ is to be minimized (as in the case of least squares or loss function), the sequence $\left\{\mathbf{F}(\widehat{\mathbf{\Theta}}^{0}), \mathbf{F}(\widehat{\mathbf{\Theta}}^{1/3}), \mathbf{F}(\widehat{\mathbf{\Theta}}^{2/3}), \mathbf{F}(\widehat{\mathbf{\Theta}}^{1}), \mathbf{F}(\widehat{\mathbf{\Theta}}^{1+1/3}), \mathbf{F}(\widehat{\mathbf{\Theta}}^{1+2/3}), \ldots\right\}$ will be monotone decreasing instead. This ensures that local convergence will always be obtained.

Despite the assurance of local convergence, global convergence is not guaranteed. Therefore, it can be useful to try different starting points to make sure that the estimates found are not suboptimal. We discuss this problem in more detail in Appendix B. The rate of convergence is heavily dependent on the chosen objective function. Although we have not studied the rate of convergence theoretically, in practice we found the algorithm to converge in few iterations.

Note that this method only works assuming that the objective function to be minimized or maximized remains the same in Step 2.1, Step 2.3 and Step 2.5. That is, one cannot estimate $\boldsymbol{\beta}$ while holding $\boldsymbol{p}$ and $\boldsymbol{q}$ constant using a mixed model with restricted maximum likelihood (REML) estimation and then estimate $\boldsymbol{p}$ while holding $(\boldsymbol{\beta}, \boldsymbol{q})$ constant using a ordinary linear model with least squares minimization or a mixed model with maximum likelihood (ML) estimation.

This can be shown, without loss of generality, by assuming that $\mathbf{F}(\widehat{\mathbf{\Theta}})$, $\mathbf{G}(\widehat{\mathbf{\Theta}})$, $\mathbf{H}(\widehat{\mathbf{\Theta}})$ are the objective functions to be maximized in Step 2.1, Step 2.3 and Step 2.5 respectively. This means that $\mathrm{F}\left(\widehat{\mathbf{\Theta}}^{\mathrm{i}}\right) \leq \mathrm{F}\left(\widehat{\mathbf{\Theta}}^{\mathrm{i}+1/3}\right)$, $\mathrm{G}\left(\widehat{\mathbf{\Theta}}^{\mathrm{i}+1/3}\right) \leq \mathrm{G}\left(\widehat{\mathbf{\Theta}}^{\mathrm{i}+2/3}\right)$ and $\mathrm{H}\left(\widehat{\mathbf{\Theta}}^{\mathrm{i}+2/3}\right) \leq \mathrm{H}\left(\widehat{\mathbf{\Theta}}^{\mathrm{i}+1}\right)$.On the other hand, $\mathrm{F}\left(\widehat{\mathbf{\Theta}}^{\mathrm{i}+1/3}\right)$ is not guaranteed to be smaller or equal to $\mathrm{F}\left(\widehat{\mathbf{\Theta}}^{\mathrm{i}+1}\right)$, thus converge of the sequence $\left\{\mathbf{F}(\widehat{\mathbf{\Theta}}^{0}), \mathbf{F}(\widehat{\mathbf{\Theta}}^{1/3}), \mathbf{F}(\widehat{\mathbf{\Theta}}^{1}), \mathbf{F}(\widehat{\mathbf{\Theta}}^{1+1/3}), \mathbf{F}(\widehat{\mathbf{\Theta}}^{2}), \ldots\right\}$ is not guaranteed. As long as the same algorithm and hyperparameters are used in all alternating parts of the algorithm (Step 2.1, Step 2.3 and Step 2.5), the model will be minimizing/maximizing the same objective function.

## Implementation

We developed an R package (*LEGIT*) and a set of SAS macros (*LEGIT_SAS*) to implement alternating optimization. The R package applies alternating optimization for generalized linear models (GLM) using the **glm** function (R Development Core Team, 2016). In addition, it also provides functions to perform cross-validation and stepwise searches for the alternating optimization models. The stepwise search function can do 'forward search', 'backward search' and 'bidirectional search' based on multiple criteria (i.e., AIC, BIC, cross-validation error, cross-validation AUC). This function can also be run in 'interactive mode'; in this mode, the user is provided with information on the best choices of variables to be added/dropped and the user can select which variable to add/drop.

For SAS, there are three macros implementing alternating optimization: (i) for generalized linear mixed models using PROC GLIMMIX; (ii) for linear mixed models using PROC MIXED; and (iii) for logistic regression using PROC LOGISTIC. In addition, we also provide macros to perform 'leave-one-out cross-validation' (LOOCV) and 'forward search' in combination with the alternating optimization models.

The main advantage of running alternating optimization in R is that it is much faster than in SAS. It takes approximately 10 to 20 seconds to perform cross-validation in R as opposed to 5 to 7 minutes in SAS. The advantage of SAS, however, is that it also implements generalized linear mixed models in addition to generalized linear models. Implementing mixed models in R is complicated by the absence of a mixed model software that enables the use of a known covariance matrix. Both the *LEGIT* package and *LEGIT_SAS* macros are available online on GitHub (github.com/AlexiaJM). Additionally, the *LEGIT* package can also be downloaded from the Comprehensive R Archive Network (CRAN).

## Results

### Simulation study

To test the performance of the alternating optimization approach, we constructed two synthetic examples in which the true model coefficients and variable distributions were known: 1) a two-way G×E and 2) a three-way G×$E_1$×$E_2$. These synthetic examples were inspired by a real example from the MAVAN study for predicting children's attentional capacity (presented in the following section: Examples from the MAVAN study) for which there were 4 genetic variants, 2 gene × gene interactions, and 3 environmental factors. The genetic variants and the environmental factors were sampled from the following distributions:

$$\boldsymbol{g}_j \sim Binomial(n = 1, p = .30),$$
$$\boldsymbol{e}_l \sim Normal(\mu = 0, \sigma = 1.5),$$

where $j = 1, 2, 3, 4$ and $l = 1, 2, 3$. In both examples, the true function for the genetic score and the environmental scores were:

$$\boldsymbol{g} = .2\boldsymbol{g}_1 + .15\boldsymbol{g}_2 - .3\boldsymbol{g}_3 + .1\boldsymbol{g}_4 + .05\boldsymbol{g}_1\boldsymbol{g}_3 + .2\boldsymbol{g}_2\boldsymbol{g}_3,$$
$$\boldsymbol{e} = -.45\boldsymbol{e}_1 + .35\boldsymbol{e}_2 + .2\boldsymbol{e}_3.$$

We looked at two scenarios, one assuming a medium effect size ($R^2$ = .30) and one assuming a small effect size ($R^2$ = .15), to test the ability of the alternating optimization method to handle noise. These scenarios are realistic as larger effect sizes are rare in gene by environmental studies; genes and environments are rarely enough to fully predict an individual physical or mental health outcome. We note that alternating optimization, being able to handle multiple genetic variants and environments with optimal weights, tends to lead to models with larger effect sizes that what is traditionally observed (See Table 3 and 4). We further set two different starting points, one assuming equal weights and one using the true weights, to test the ability of the model to perform when an incorrect starting point is used compared to an optimal starting point.

For example 1, the function defining the relationship between the genetic score and the environmental score with the outcome was:

$$\boldsymbol{y} = 5 + 2\boldsymbol{g} + 3\boldsymbol{e} + 4\boldsymbol{ge} + \boldsymbol{\varepsilon},$$

where
$\boldsymbol{\varepsilon} \sim Normal(\mu = 0, \sigma = 4.36)$ for the medium effect size scenario ($R^2$ = .30) and
$\boldsymbol{\varepsilon} \sim Normal(\mu = 0, \sigma = 6.78)$ for the small effect size scenario ($R^2$ = .15).

For example 2, the function defining the relationship between the genetic score, the environmental score and the additional environmental factor with the outcome was:

$$\boldsymbol{y} = 5 + 2\boldsymbol{g} + 3\boldsymbol{e} + \boldsymbol{z} + 5\boldsymbol{ge} + 1.5\boldsymbol{ez} + 2\boldsymbol{gz} + 2\boldsymbol{gez} + \boldsymbol{\varepsilon},$$

where
$\boldsymbol{\varepsilon} \sim Normal(\mu = 0, \sigma = 12.31)$ for the medium effect size scenario ($R^2$ = .30),
$\boldsymbol{\varepsilon} \sim Normal(\mu = 0, \sigma = 19.19)$ for the small effect size scenario ($R^2$ = .15) and
$\boldsymbol{z} \sim Normal(\mu = 3, \sigma = 1)$.

We fit the models to "training" samples of 250, 1000 and 5000 observations respectively and verified the model predictions on a "validation" sample of 100 additional observations not used for parameter estimation. To evaluate performance, we divided the $R^2$ in the validation sample by the largest obtainable $R^2$ (around .30 in the medium effect size scenario and around .15 in the small effect size scenario). These ratios are all ≤ 1.0, an optimal model would have a value of 1.0. Confidence interval coverage (95%) of the genetic, environmental, and main model coefficients were also examined. We ran 100 simulations for each scenario to ensure that results were unbiased and evaluation criteria were averaged across all simulations.

The results of the simulations are presented in Table 2. The simulations with N=1000 and N=5000 obtained very large $R^2_{val}$ / $R^2_{max}$ (.9 to 1), on both examples and effect sizes. Furthermore, in terms of $R^2_{val}$ / $R^2_{max}$, the N=250 sample obtained .87 on example 1 with medium effect sizes, .64 on example 1 with small effect size, .66-.68 on example 2 with medium effect size and .07-.08 on example 3 with small effect size.

Depending on the effect size (.15 or .30) and complexity of the model (2-Way or 3-Way), we found that $Genes_{cov}$ = .70-.87, $Env_{cov}$ = .92-.98 and $Main_{cov}$ = .77-.82 for the small sample size (N=250), $Genes_{cov}$ = .85-.92, $Env_{cov}$ = .96-.98 and $Main_{cov}$ = .81-.91 for the moderate sample size (N=1000), and $Genes_{cov}$ = .90-.93, $Env_{cov}$ = .97-.98 and $Main_{cov}$ = .87-.90 for the large sample size (N=5000). Assuming equal weights as the starting point resulted in similar validated $R^2$ and coverage estimates as when starting with the true weights as starting point.

## Examples from the MAVAN study

To further examine the performance of the alternating optimization approach, we used real data from the Maternal Adversity, Vulnerability, and Neurodevelopment (MAVAN) study. The first example applies a two-way G×E longitudinal mixed model to predict a continuous outcome, namely negative emotionality measured at 3, 6, 18 and 36 months. The second example uses a three-way G×$E_1$×$E_2$ longitudinal mixed model to predict a continuous score of attentional competence measured at 18 and 24 months. Details on how to adapt alternating optimization for mixed models are available in Appendix C.

**Sample description.** For both examples, we used mother-child dyads from the ongoing longitudinal Maternal Adversity, Vulnerability, and Neurodevelopment (MAVAN) project. The MAVAN is a Canadian community-based cohort of 627 women recruited during pregnancy in Montreal (Qc.) and Hamilton (On.). Women were recruited in maternity hospitals from 2003 to 2009 during their routine ultrasound examinations. To be eligible, women had to be 18 years of age or over at the expected date of delivery, with a singleton and full-term pregnancy (≥37 weeks). Exclusion criteria were the presence of severe chronic maternal illness, past obstetrical complications or major fetal/infant anomaly. The average age of women at recruitment was 30.3 years.

The mothers were interviewed between 24 and 36 weeks of pregnancy and the dyads were assessed at 3, 6, 12, and 18 months and yearly from 24 months onwards. Maternal health and well-being were assessed each year using validated measures of maternal mental health, social and family functioning and socio-economic status. The children were assessed with age-appropriate measures of temperament, socio-emotional development and psychopathology. A detailed description of recruitment, procedure, and measures has been published elsewhere (O'Donnell et al., 2014). Retention rates for the MAVAN are 97.4% at 6 months, 84.0% at 18 months, and 80.6% at 36 months.

**Evaluation criteria.** To assess the quality of model fit, we focused on three evaluation criteria: (i) Akaike information criterion (AIC), (ii) Bayesian information criterion (BIC), (iii) in-sample effect size and (iii) out-of-sample effect size. Note that the genetic and environmental scores parameters were accounted for in the AIC and BIC. The in-sample effect size was defined as the regular $R^2$ and the out-of-sample effect size was defined as the leave-one-out cross-validated (LOOCV) $R^2$, which can be defined in the following way for repeated measures:

$$R^2_{LOOCV} = 1 - \frac{\sum_{i=1}^{n_{subject}} \sum_{t=1}^{n_i} \left(y_{it} - \hat{y}_{(i)t}\right)^2}{\sum_{i=1}^{n_{subject}} \sum_{t=1}^{n_i} \left(y_{it} - \bar{y}\right)^2},$$

where $n_{subject}$ is the number of subjects (mother-child pairs in MAVAN analyses), $n_i$ is the number of available time-points for the $i^{th}$ subject (e.g. with ITSEA, a child can have either 18 months, 24 months or both time points) $y_{it}$ is the $i^{th}$ outcome variable at time-point t, $\bar{y}$ is the average of the outcome variable and $\widehat{y_{(i)t}}$ is the prediction of $y_{it}$ using the model fitted without $y_{i1}, \dots, y_{in_i}$.

**Predicting Negative emotionality at 3, 6, 18 and 36 months.** Negative emotionality (NE) is a temperamental dimension that reflects a generally stable tendency of the child toward increased emotional reactivity with regards to negative situations, such as anger, fear or sadness (Lemery et al., 1999). In previous studies, we showed that prenatal maternal depression interacts with a multi-locus genetic score to predict negative emotionality from 3 to 36 months (Green et al., 2016) and that mother's own traumatic childhood experiences interact with offspring 5-HTTLPR genotype in predicting NE at 18 and 36 months (Bouvette-Turcot et al., 2015). We derived NE using the Infant Behavior Questionnaire (IBQ) at 3 and 6 months (Gartstein and Rothbart, 2003) and the Early Childhood Behavior Questionnaire (ECBQ) at 18 and 36 months (Putnam et al., 2006). The multi-locus score was composed of two genetic variants assuming equal weights:

1) The 48bp vntr in exon 3 of the DRD4 gene. The value for this genetic variant (DRD4) was set to 1 when the child possessed 6 or more repeats and to 0 otherwise (Auerbach et al., 1999).
2) The 43bp vntr in the promoter region of 5-HTT coupled with its transcriptional efficiency, based on the polymorphism in rs25531 (Hu et al., 2006). The value for this genetic variant (5-HTTLPR) was set to 0 when the child was $L_AL_A$ and to 1 otherwise (Pluess et al., 2011).

Our aim was to improve this model by using alternating optimization to determine the optimal weights of DRD4 and 5-HTTLPR and to explore whether adding additional genetic variants would improve the predictive ability of the genetic score. To this end, we included the oxytocin peptide gene (OXT), as previous work from our group found that polymorphisms in this gene were closely linked to aspects of maternal care (Mileva-Seitz et al., 2013), while others reported significant interactions between oxytocinergic genes and 5-HTTLPR in predicting child NE (Montag et al., 2011). We did not use the same covariates as in Green et al. (2016) but retained only those that contributed to the out-of-sample effect size (i.e., postnatal maternal depression at the previous time point, maternal college education, Material/social deprivation index and mother's age at birth) in order to prevent any model bias. Pre- and postnatal maternal depression was assessed using the Center for Epidemiologic Studies Depression Scale (CES-D) self-report questionnaire (Radloff, 1977), while the Material/Social deprivation index (Pampalon et al., 2009) was constructed from census data and transformed into quintiles with higher values representing lower SES. We used separate intercepts for NE at 3 and 6 months and at 18 and 36 months, as these were measured using different instruments (i.e., IBQ and ECBQ, respectively). The model fitted was a two-way longitudinal mixed model with a continuous outcome. After removing participants with incomplete data and 5 outliers defined as having LOOCV standardized residuals > 2.79, the final sample size was N=607. The following three models were fitted:

1) **Baseline model**: including covariates only.
2) **2-way model**: a two-way interaction model including prenatal maternal depression and a genetic score of DRD4 and 5-HTTLPR, assuming equal weights for both gene variants.

3) **Alternating optimization for genetic score**: a two-way interaction model including prenatal maternal depression and a genetic score comprising DRD4, 5-HTTLPR, assuming unequal genetic weights.
4) **Alternating optimization with additional OXT**: a two-way interaction model including prenatal maternal depression and a genetic score comprising DRD4, 5-HTTLPR, and OXT, assuming unequal genetic weights.

Model results are shown in Table 3. The two-way interaction model with equal weights was a better fit than the covariates only model (the in-sample $R^2$ increased from .11 to .17, the out-of-sample effect size increased from .08 to .13, the AIC decreased from 921.6 to 901.7 and the BIC decreased from 974.57 to 964.56).

Estimating the weights of the genetic score of the previous model did not meaningfully improve model fit (in-sample $R^2$ increased from .17 to .18, the out of sample $R^2$ did not change, the AIC increased from 901.7 to 902.48 and the BIC from 964.6 to 968.4). This can be explained by the fact that the equal weights were very similar to the optimal weights. This issue is further addressed in the Discussion.

We found that a single nucleotide polymorphism (rs2740210) in the oxytocin peptide gene (OXT) contributed meaningfully to the genetic score (coded as 1 = AC or AA genotypes, 0 = CC genotype) (Jonas et al., 2013). When including OXT in the model, all model fit parameters improved (in-sample $R^2$ increased from .17 to .20, the out-of-sample effect size increased from .13 to .14, the AIC decreased from 901.7 to 892.5 and the BIC from 968.39 to 954.22). In conclusion, the final model using alternating optimization provided the best fit to describe our data. We estimated the relative contribution of each genetic variant and found that DRD4 contributed 19% ($p = .10$), 5-HTTLPR contributed 44% ($p < .0001$) and OXT contributed 37% ($p < .0001$) to the genetic score, all with a positive directionality. The interaction from the final model is illustrated in Figure 2.

Importantly, using alternating optimization, we found suggestive evidence of an association between the oxytocin peptide gene and NE. A previous study by Montag et al. (2011) also observed a link between 5-HTTLPR, oxytocin, and NE, although they reported an interaction between 5-HTTLPR and the oxytocin gene, which we did not replicate in the present analysis. One explanation for this might be the difference in the oxytocin gene studied: we examined the oxytocin peptide gene, while Montag et al. (2011) studied the oxytocin receptor gene.

In our original study on the prediction of NE (Green et al., 2016), we found that the G×E was consistent with the differential susceptibility model at 3 and 6 months and the diathesis-stress model at 36 months (Belsky, 1997a, Belsky, 1997b). In the present study, we were unable to test the specific form of the interaction since our alternating optimization approach has not been yet adapted to work with the confirmatory modelling approach by Belsky et al. (2013). A more detailed analysis and discussion of the possible mechanisms are forthcoming from our group.

**Predicting attention at 18 and 24 months.** Attentional functioning was obtained from the competence domain of the Infant-Toddler Social and Emotional Assessment (ITSEA) (Briggs-Gowan and Carter, 1998, Briggs-Gowan and Carter, 2007). The ITSEA was administered at 18 and 24 months. We constructed a $G \times E_1 \times E_2$ model to represent the interactions between a genetic score (G) of dopaminergic genes, prenatal maternal depression ($E_1$), and a maternal sensitivity score ($E_2$) consisting of early postnatal maternal behavior. Microanalytic measures of maternal behavior were extracted from a videotaped session of 20 min of nonfeeding interaction followed by a 10 min divided attention maternal task at 6 months. The BEST coding system (Educational Consulting, Inc. Florida, US; S & K NorPark Computer Design, Toronto) was used to generate duration and frequency data for multiple maternal behaviors by use of a computer keyboard with keys indexed for each behavior. The percentage was subsequently coded as the duration of each behavior divided by the total duration (minus talking to someone else and feeding). Inter-rater reliability was obtained by having two observers code the same 18 videos of mother-infant interactions. Inter-rater reliability was high, with r values of 0.74 to 0.90 for looking away frequency and duration, respectively.

Before the alternating optimization approach, to construct a $G \times E_1 \times E_2$ model, we only used a single genetic variant and two environments at a time. The percentage of time mothers spent looking away from their infant was the only measure we used of early postnatal maternal behavior ($E_2$) and we ran separate models for the 5 following genetic variants:

1) An SNP of the DRD2 gene, rs1800497. The variable for this genetic variant (BDNF) was set to 1 when A2/A2 and 0 otherwise (Holmboe et al., 2010).
2) The 48bp vntr in exon 3 of the DRD4 gene. The variable for this genetic variant (DRD4) was set to 1 when the child possessed 6 or more repeats and to 0 otherwise (Schmidt et al., 2001).
3) The 40bp vntr in the 3' region of the DAT gene. The variable for this genetic variant (DAT1) as set to 1 when 10/10 and 0 otherwise (Holmboe et al., 2010).
4) An SNP of the BDNF gene, rs6265. The variable for this genetic variant (BDNF) was set to 1 when Val/Val and 0 otherwise (Lang et al., 2009).
5) An SNP of the COMT gene, rs4680. The variable for this genetic variant (COMT) was set to 1 when Met/Met and 0 otherwise (Holmboe et al., 2010).

We found that all models, except the one with DRD2, followed a similar pattern, in terms of the coefficients and the plots. Accordingly, we created a genetic score, assuming equal weight, using DRD4, DAT, BDNF, and COMT. Still, it remained unclear how much each gene variants were really contributing since assuming equal weight is overly simplistic. We also wondered whether there could be gene × gene interactions and others aspects of maternal behavior that could contribute additionally to maternal sensitivity ($E_2$).

Using alternating optimization, we aimed to (i) determine the correct weights of the four genetic variants, (ii) detect the presence of gene × gene interactions within the genetic score and (iii) identify additional aspects of maternal behavior that significantly contribute to the environmental score. Using a forward stepwise selection, we added the gene × gene interactions (6 possible combinations) and maternal behaviors (11 frequency/percentage behaviors) one-by-one into the scores and retained only those which increased the out-of-sample effect size.

Similarly, we included only the covariates that contributed to the out-of-sample effect size (i.e., postnatal maternal depression at 12 months and child self-regulation at 6 months). The prenatal and postnatal depression variables were constructed from the CES-D self-report questionnaire (Radloff, 1977) and self-regulation at 6 months was measured using the IBQ (Gartstein and Rothbart, 2003). We included separate intercepts for attentional competence 18 months and 24 months to adjust for the fact that baseline attention is significantly higher at 24 months ($p < .0001$). The model fitted was a two-way longitudinal mixed model with a continuous outcome. After removing participants with incomplete data and 5 outliers with LOOCV standardized residuals greater than 2.73, the sample size for the final analysis was N=212. The following four models were fitted:

1) **Baseline model**: including covariates only.
2) **3-way model**: a three-way interaction model including prenatal maternal depression, maternal sensitivity (defined as looking away percentage) and a dopaminergic genetic score (composed of DRD4, DAT1, BDNF, and COMT), assuming equal weights for all genetic variants.
3) **Alternating optimization for environmental score with additional variables**: a three-way interaction model including prenatal maternal depression, a maternal sensitivity score (consisting of looking away percentage, percentage of time spent with kissing and frequency of physical play without toys) and a dopaminergic genetic score (composed of DRD4, DAT1, BDNF, and COMT), assuming equal genetic weights.
4) **Alternating optimization for genetic score**: an extension of Model 3 by not assuming equal weights for the dopaminergic genetic score.
5) **Alternating optimization with additional G×G**: an extension of Model 4 by including DRD4×DAT1 and DRD4×COMT gene × gene interactions into the dopaminergic genetic score.

Model results are shown in Table 4. The three-way interaction model with equal weights proved to be a better fit than the baseline model (in-sample $R^2$ increased from .12 to .23, out-of-sample $R^2$ increased from .08 to .12 and AIC decreased from 242.6 to 235). However, the BIC was higher (from 262.07 to 274) due to the very large penalty on additional parameters.

Adding maternal sensitivity variables and estimating their weights further improved the model fit (in-sample $R^2$ increased from .23 to .35, out-of-sample $R^2$ increased from .12 to .24 and AIC decreased from 234.97 to 213.11). In this case, the model was better than the baseline model with regards to the BIC (from 262.07 to 256.51) even though the number of parameters was important (13).

Estimating the weights of the genetic score did not meaningfully improve the model fit (in-sample $R^2$ increased from .35 to .38), thus the AIC and BIC increased (the AIC from 213.11 to 213.39 and the BIC from 256.51 to 260.05). Furthermore, the out of sample $R^2$ decreased slightly (from .24 to .20). This is similar to what we observed in the previous MAVAN example on the prediction of NE. This can be explained by the fact that equal weights were close enough to the optimal weights. This issue is addressed in more detail in the Discussion.

In the final step, we added the G×G effects and observed an improvement of the model fit (in-sample $R^2$ increased from .38 to .41, out-of-sample $R^2$ increased from .20 to .25, AIC decreased from 213.39 to 208.28 and BIC decreased from 260.05 to 255.64). This model was the best model in all aspects. Thus, as a result of the stepwise inclusion of variables into the genetic score and maternal sensitivity score, we were able to identify gene × gene interactions and additional important aspects of the mother-child interaction (i.e., play, tactile stimulation). The interaction effect from the final model is shown in Figure 3.

In the final model, we estimated the following relative contributions of the genetic variants: DRD4 contributed 12% positively ($p = .04$), DAT1 contributed 26% positively ($p < .0001$), BDNF contributed 5% positively ($p < .0001$), COMT contributed 5% positively ($p < .0001$), DRD4 × DAT1 contributed 20% negatively ($p = .004$) and a DRD4 × COMT contributed 32% positively ($p < .0001$). With respect to maternal sensitivity, we estimated the following relative contributions of the comprising parameters: maternal looking away percentage (inattention aspect) contributed 45% negatively ($p < .0001$), frequency of physical play without toys (play aspect) contributed 34% positively ($p < .0001$) and the percentage of kissing (tactile aspect) contributed 21% positively ($p < .0001$).

A more detailed analysis and discussion of the possible mechanisms are forthcoming from our group. Future work in our group will study the expansion of the multilocus dopaminergic gene score by adding additional dopaminergic genes which have been associated with aspects of observed maternal behavior in a previous MAVAN study (Mileva-Seitz et al., 2012) and investigating further the interaction.

## Discussion

We presented a novel approach called alternating optimization, with implementation in R and SAS, to estimate genetic and environmental scores when interactions between the scores are of interest. We demonstrated how to construct a G×E model using alternating optimization and presented a simple argument showing that the model converges toward a local optimum. We then showed that, using two synthetic examples with known coefficients, the alternating optimization approach performed well under varying sample sizes and effect sizes. Furthermore, we have shown that incorrect starting points did not have any noticeable impact on the results; this suggests that global convergence was likely achieved. We then illustrated the utility of our model using real data from the MAVAN study to predict childhood negative emotionality and attention problems. Both the synthetic as well as the real-life examples confirmed the validity and power of the alternating optimization approach for dealing with multiple genetic and/or environmental variants and their complex interactions.

### Advantages

There are many advantages of using this approach for the construction of models with multiple genetic variants, environments, and potential interactions. First, we need to estimate significantly fewer parameters relative to traditional methods, where the interaction effect of each individual genetic variant with each environmental exposure must be estimated (at least before variable selection techniques are applied). With alternating optimization, a G×E model has 3 main model parameters and a $G \times E_1 \times E_2$ has 7 main model parameters; in addition, each score (genetic or environmental) contains $n_s$ parameters but only $(n_s - 1)$ parameters have to be estimated, where $n_s$ is the number of elements in the score. This is because one weight of the score is always fixed (except for the sign) by the constraint that the sum of the absolute weights must equal to 1, e.g., $p_1 = \pm\left(1 - \sum_{j=2}^{k}|p_j|\right)$ and $q_1 = \pm\left(1 - \sum_{l=2}^{s}|q_j|\right)$. Assuming that a model had 4 genetic variants, 2 gene x gene interactions, 3 environments of one category and another environment that were interacting in a $G \times E_1 \times E_2$; in an alternating optimization model we would have 18 parameters to estimate but in a standard interaction model, we would have 59 parameters to estimate. This is shown clearly in Figure 1; for three-way interaction models, alternating optimization scales linearly with the number of elements $O(k + s + l)$ while traditional methods scale in cubic time $O(ksl)$. The more interactions the main model contains, the stronger the difference between the number of parameters in the methods is. The way some modern methods deal with this problem is by assuming sparsity, i.e. most interaction terms are assumed to be zero so very few have to be estimated. This assumption does not need to be made with our approach.

Second, we are assigning a biologically plausible structure to the model. Many of the modern methods for G×E interaction modeling do not force a specific model structure but learn it automatically, which gives them more flexibility at the expense of a sound biological structure. With alternating optimization, we need to select the proper model structure for how the genetic and environmental scores are thought to interact with one another. Note that this can be seen as both a disadvantage because it's not fully automatized and flexible, and an advantage because we can select a plausible model structure based on a priori understanding of the genes and environments involved.

Third, the alternating optimization is very fast and convergence is obtained rapidly in few iterations. Fourth, this approach is guaranteed to converge to a locally optimal solution and as we have shown, it tends to converge very closely to the true solution.

**Disadvantages or Limitations**

Simulations from Table 2 showed low predictive power in the small sample and small effect size scenario. This is accounted for by the categorical nature of the genetic variants. Replacing the genetic variants by normally distributed variables with the same mean and variance led to ratios of $R^2_{val}$ / $R^2_{max}$ equal to .81 and .83 for the equal weights and true weights starting points respectively in example 2 with small effect size (instead of .07 and 08 as in Table 2).

Confidence interval coverage should be near 95% if estimates are unbiased. The simulations from Table 2 show that, in multiple cases, the coverage values were somewhat lower than expected (.95). We found the lower coverage in genes to be partially explained by the categorical nature of the genetic variants and the gene by gene interactions. Additionally, we speculate that the relatively lower coverage for the main model parameters might be due to their dependence on latent variables rather than observed variables.

Considering the minimal improvement in model fit when estimating the genetic variants weights in the two examples from the MAVAN dataset (see Table 3 and 4), one might conclude that alternating optimization is not helpful, and that equal weights would be preferable. However, it is important to note that:

1) The further away the true parameters are from equal weights, the larger the potential benefit of using alternating optimization. Given that in both examples from the MAVAN study, the true weights were close enough to equal thus estimating the weights did not significantly improve model fit.
2) In practice, the various genetic and environmental effects on a phenotype are rarely uniform in direction and magnitude. The purpose of alternating optimization is to estimate both thereby allowing unanticipated relationships between the manifest variables comprising each latent score.
3) We rarely know beforehand the complete list of variables that need to be included in an explanatory model of complex psychological outcomes. Rather, we often have only a few potential candidates. In the example we presented on of predicting toddlers' attentional competence, due to including the manifest variables into the construction of the latent variables (G and $E_2$) in a stepwise manner, we identified gene × gene interactions and additional important aspects of the mother-child relationship (i.e., play, tactile stimulation). These processes might have gone unnoticed without using alternating optimization.

Alternating optimization still has some disadvantages. First, it is not guaranteed to converge to the global optimum thus it could be sometimes necessary to try different starting points. In practice, alternating optimization led to solutions really close to the true solutions but this might not always be true, especially with a lot of genetic variants and or environments. Second, although it scales linearly in terms of the number of parameters to estimates, variable selection cannot be done with standard regularization techniques because the full model is never seen. Therefore, stepwise search (using the AIC/BIC or cross-validation error as a guideline) needs to be used instead which is slower and not guaranteed to lead the optimal set of variables.

One limitation of the present study is that we have not examined mathematically the convergence properties of the method. Thus, we cannot provide information regarding the convergence rate, although we found convergence to be rather quick in practice. Future work should explore mathematical convergence properties of the model in more detail.

Furthermore, we only verified how far the local solutions were from the global solutions in specific examples with a small number of genetic variants and environments. When including a large number of genetic variants and/or environments, convergence to an optimal solution might be more difficult to attain than in the cases presented here. This could be investigated with simulations and real examples from larger datasets.

We also note that there might be more efficient and optimal ways to perform variable selection than the currently suggested stepwise approach. For instance, it might be possible to devise a form of regularization that could be used with alternating optimization. More research on variable selection with alternating optimization needs to be done.

## Relationship with other conceptual models for G×E

In addition to modelling G×E, one might also want to test specific hypotheses regarding the form of the interaction. The two most relevant theories are the diathesis-stress model (Zubin and Spring, 1977) and the differential susceptibility model (Belsky, 1997a, Belsky, 1997b). The diathesis-stress model assumes that a negative environment acts as a risk factor, while the differential susceptibility model assumes that a positive environment leads to a good outcome and a negative environment leads to a bad outcome. A few approaches have been made to distinguish diathesis-stress from differential susceptibility (Belsky et al., 2013, Roisman et al., 2012). Currently, alternating optimization does not test for the different types of interactions, but it would be possible to adapt the approach by Belsky et al. (2013) to work within a latent variable model relying on alternating optimization. This is something we would like to do in the future.

Poor measurement scaling can cause interactions to appear significant when there is actually no interaction (or vice-versa). This is generally due to monotonic non-linear transformations of the data and non-normality of the outcome/residuals (Molenaar and Dolan, 2014). There is no clear-cut solution to poor measurement scaling however researchers should always make sure that the model they use is adequate for the data and that no unnecessary transformations are used. Considering that alternating optimization can be used with most models and with any link function (e.g., logit, probit, etc.), measurement scaling can and should be accounted for by choosing the most appropriate model for one's data.

## Variations

Although we presented the method as a linear regression model with a two-way interaction, we can adapt this method to other model variants. We review some of the possibilities below.

To fully characterize the relationship between the covariates and a complex trait or disease, it is reasonable to assume that one would need something more complex than a two-way interaction model. As an example, in developmental psychopathology, it is thought that the influence of genetic variants on child outcome is not only moderated by the prenatal environment (e.g., prenatal programming by intra-uterine growth retardation or maternal distress (Pluess and Belsky, 2011, Schlotz and Phillips, 2009) but also by the postnatal environment (maternal care (Meaney, 2001) or socioeconomic status (McLoyd, 1998)). The prenatal and the postnatal environments of the child are thought to have their own influence on the outcome while dependent on one another. Therefore, a three-way interaction is necessary to fully capture this mechanism. The alternating optimization approach can easily be adapted for three-way interaction models, more details are available in Appendix C.

The alternating optimization approach has been presented in the context of a standard linear model which assumes, by definition, that the outcome is continuous. However, phenotypic outcomes can often be binary or categorical. Similarly, the approach has only been presented assuming fixed-effects, but mixed models are frequently required to account for siblings, site, ethnicity, genetic ancestry or to model an outcome with repeated measures. To be able to use non-continuous outcomes and random-effects, the approach has been adapted for generalized linear mixed models (GLMM). Full details are available in Appendix C.

A genetic (or environmental) score is just a weighted sum of genetic variants (or environments) but it is also possible to include interactions within the genetic (or environmental) score. For example, a genetic score that contains two-way interactions, assuming that we include all interactions terms (generally unnecessary), would look like this:

$$\boldsymbol{g} = \sum_{j=1}^{k} p_j \boldsymbol{g}_j + \sum_{j_1=1}^{k} \sum_{j_2=1}^{k} I(j_1 \neq j_2) p_{j_1 j_2} \boldsymbol{g}_{j_1} \boldsymbol{g}_{j_2} .$$

Given the important power demands of four-way and five-way interaction models ($2^K - 1$ variables needed for a $K$-way interaction model), the ability to construct a G×E or G×$E_1$×$E_2$ model where the genetic and/or environmental scores each contain two-way or three-way interactions means that we can create models with effectively four-way or five-way interactions with many fewer parameters than normally required.

This permits the modelling of complex cell signaling systems in accord with evolving evidence of G×G. For example, in the dopaminergic system, there are important findings of DAT × COMT (Dreher et al., 2009, Prata et al., 2009, Yacubian et al., 2007), DRD4 × BDNF (Kaplan et al., 2008), and DRD2 × DRD4 interactions (Beaver et al., 2007). G×G are consistent with the curvilinear functions underlying dopamine signaling and reward circuitry activation (Bigos et al., 2016). Considering that little or too much dopamine in the brain might lead to suboptimal neural activation, such a model captures the possibility that a beneficial genetic variant associated with increased dopamine might become a risk factor when combined with one or more dopamine-increasing genetic variants.

Although we presented the model has having only one genetic score and one environmental score, nothing prevents the creation of multiple scores. We could create more than one genetic score, for example, one for every system (dopamine, serotonin, neuronal growth factors, etc.) and more than one environmental score, for example, one for every developmental period (prenatal, early life, childhood, etc.).

It can also be noticed that the model we present resembles a neural network and, in fact, can be interpreted as a single-layer neural network with two nodes and identity activation function. This suggests the possibility of generalizing this approach to two or more hidden layers. For example, the first layer could be the genetic scores for different systems and the environmental scores for the different developmental periods and the second layer could be a global genetic score and a global environmental score. Nonlinear activation functions could also be used. One would need a large amount of data to be able to fit such models but it might become possible in the future with the advent of big data. Using such a complex structure could lead to very powerful predictions.

## Conclusion

To conclude, we believe that alternating optimization will aid researchers in their endeavor to simultaneously consider multiple genes and multiple environmental factors when studying important developmental and health outcomes, rather than only focusing on a single candidate gene by environment interaction. Furthermore, the reproducibility of research findings could be significantly improved by focusing on models with strong effects, accounting for multiple genetic variants, multiple environments and gene × gene or environment × environment interactions.

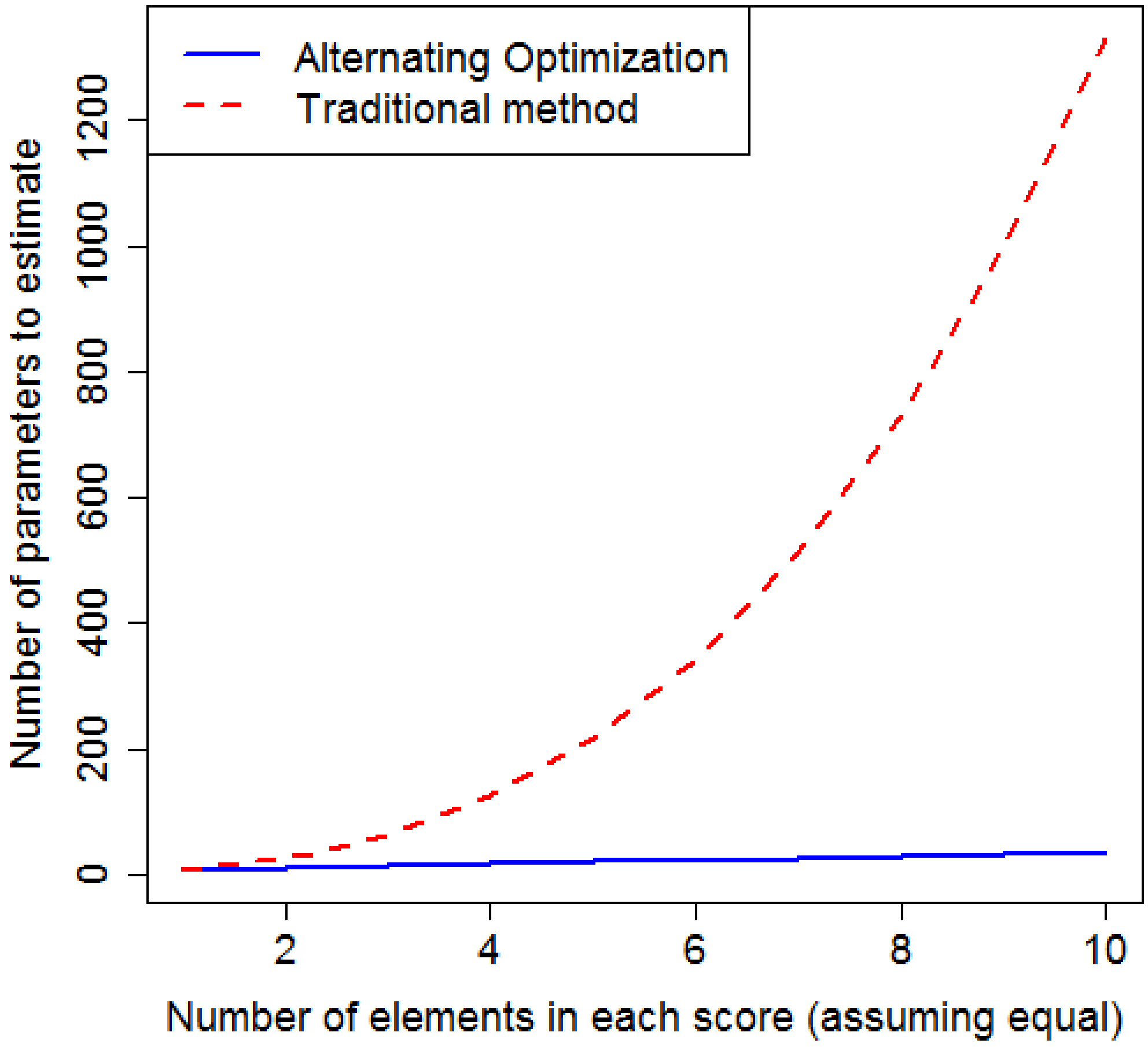

**Figure 1:** *Number of parameters to estimate as a function of the number of elements (main effects terms and gene × gene or environment × environment interaction terms) for the estimation of a G×E×$E_2$ three-way interaction model comparing alternating optimization to traditional methods.*

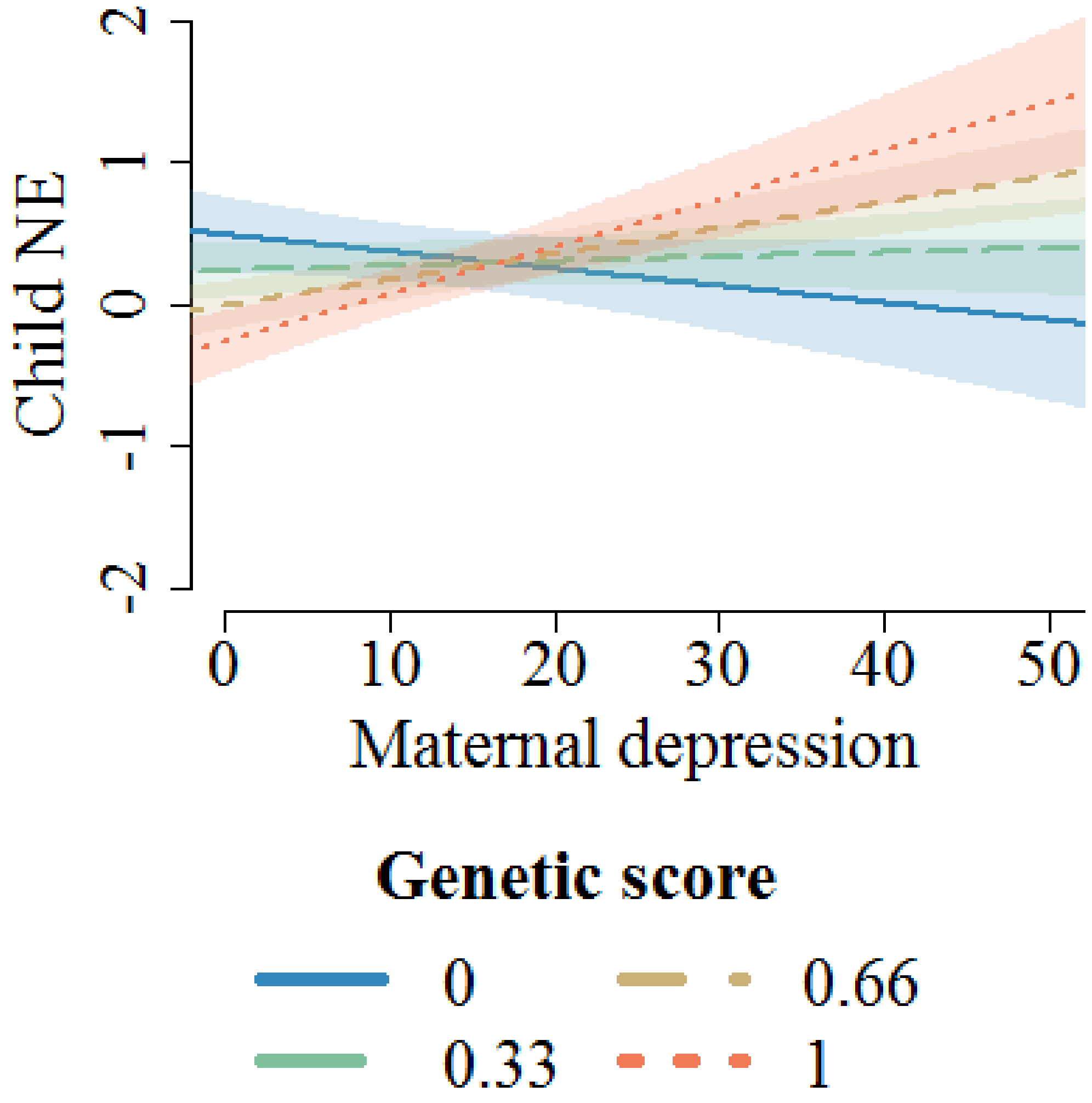

**Figure 2:** *The interaction effect of prenatal maternal depression and a multi-locus genetic score (of 5-HTTLPR, DRD4, OXT) on offspring negative emotionality (NE) at 3, 6, 18 and 36 months (longitudinal mixed model analysis with a continuous outcome, N=607). A higher multi-locus genetic score in the offspring is associated with increased early negative emotionality when the mother experiences more depressive symptoms throughout pregnancy.*

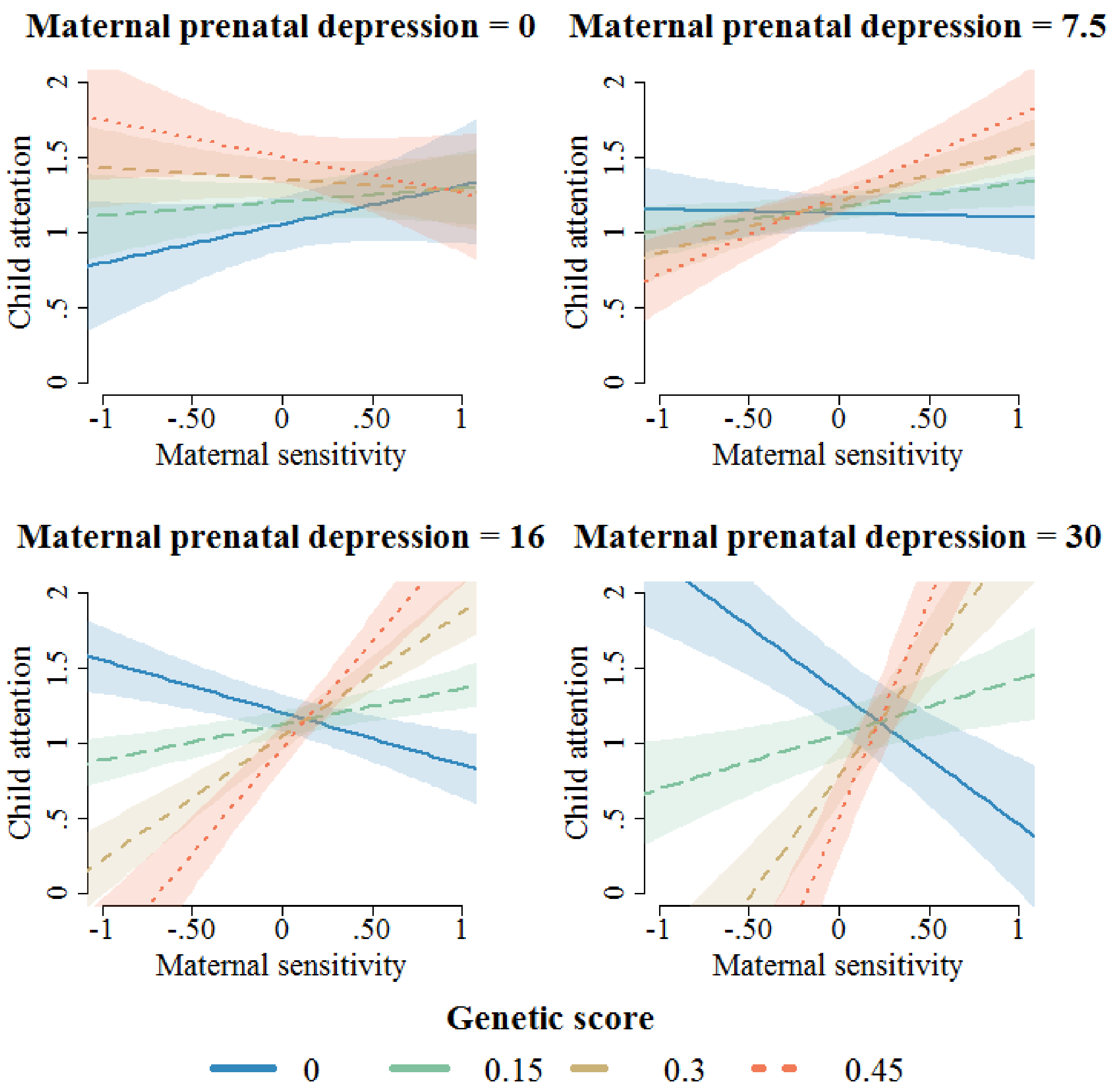

***Figure 3:*** *The interacting effects of prenatal maternal depression, maternal sensitivity and a multi-locus dopaminergic genetic score (DRD4, DAT1, BDNF, COMT, DRD4xDAT1, DRD4xCOMT) on attentional competence from 18 to 24 months (longitudinal mixed model analysis with a continuous outcome, N=212). When mothers are more sensitive, children are less influenced by their genetic scores and have better attention skills unless the mother is severely depressed during the pregnancy (in which case higher genetic score is linked to higher attention). On the other hand, when mothers are less sensitive, the influence of offspring genetic score and maternal depression become more prominent on the child attention. Overall, this suggests that the offspring's genes might be beneficial or detrimental depending on the environment.*

**Table 1:** *Number of parameters to estimate, assuming we include all terms, using the traditional way and using alternating optimization with genetic and environment scores*

| *Number of parameters to estimate* | G×E (Traditional) | G×E×$E_2$ (Traditional) | **G×E (Alternating optimization)** | **G×E×$E_2$ (Alternating optimization)** |
|---|---|---|---|---|
| Without G×G interactions | 1 + k + s + ks | 1 + k + s + r + ks + kr + sr + krs | 4 + (k-1) + (s-1) | 8 + (k-1) + (s-1) + (r-1) |
| **With G×G interactions** | 1 + k + s + ks + kk | 1 + k + s + r + ks + kr + sr + krs + kk | 4 + (k-1) + $(k-1)^2$ + (s-1) | 8 + (k-1) + $(k-1)^2$ + (s-1) + (r-1) |

*Notes*: k is the number of genetic variants (G), s the number of environments (E), r the number of environments of the second type ($E_2$). Assuming one intercept and no additional covariates.

**Table 2:** *Average (validated $R^2$ / maximum possible $R^2$) and coverage of the genetic, environmental and main model parameters for 2 examples with different scenarios and starting points, using 250 or 1000 "training" observations and 100 "validation" observations, over 100 simulations.*

| *Scenarios* / *Starting points* / *Simulation* | Medium effect size<br><br>$R^2$ = .30 with optimal parameters and infinite sample size | | Small effect size<br><br>$R^2$ = .15 with optimal parameters and infinite sample size | |
|---|---|---|---|---|
| | Equal weights | True weights | Equal weights | True weights |
| Small sample (N = 250) | | | | |
| **Example 1:**<br>*2 Way interaction model*<br>4 genes, 2 gene × gene interactions, 3 environments | $R^2_{val}$ / $R^2_{max}$ = .87<br>$Genes_{cov}$ = .87<br>$Env_{cov}$ = .98<br>$Main_{cov}$ = .82 | $R^2_{val}$ / $R^2_{max}$ = .87<br>$Genes_{cov}$ = .87<br>$Env_{cov}$ = .98<br>$Main_{cov}$ = .83 | $R^2_{val}$ / $R^2_{max}$ = .64<br>$Genes_{cov}$ = .82<br>$Env_{cov}$ = .94<br>$Main_{cov}$ = .75 | $R^2_{val}$ / $R^2_{max}$ = .64<br>$Genes_{cov}$ = .82<br>$Env_{cov}$ = .95<br>$Main_{cov}$ = .77 |
| **Example 2**:<br>*3 Way interaction model*<br>4 genes, 2 gene × gene interaction, 3 environments | $R^2_{val}$ / $R^2_{max}$ = .68<br>$Genes_{cov}$ = .74<br>$Env_{cov}$ = .96<br>$Main_{cov}$ = .81 | $R^2_{val}$ / $R^2_{max}$ = .66<br>$Genes_{cov}$ = .75<br>$Env_{cov}$ = .96<br>$Main_{cov}$ = .82 | $R^2_{val}$ / $R^2_{max}$ = .07<br>$Genes_{cov}$ = .70<br>$Env_{cov}$ = .92<br>$Main_{cov}$ = .77 | $R^2_{val}$ / $R^2_{max}$ = .08<br>$Genes_{cov}$ = .70<br>$Env_{cov}$ = .94<br>$Main_{cov}$ = .77 |
| Moderate sample (N = 1000) | | | | |
| **Example 1**:<br>*2 Way interaction model*<br>4 genes, 2 gene × gene interactions, 3 environments | $R^2_{val}$ / $R^2_{max}$ = .98<br>$Genes_{cov}$ = .92<br>$Env_{cov}$ = .98<br>$Main_{cov}$ = .84 | $R^2_{val}$ / $R^2_{max}$ = .98<br>$Genes_{cov}$ = .92<br>$Env_{cov}$ = .98<br>$Main_{cov}$ = .85 | $R^2_{val}$ / $R^2_{max}$ = .95<br>$Genes_{cov}$ = .88<br>$Env_{cov}$ = .98<br>$Main_{cov}$ = .81 | $R^2_{val}$ / $R^2_{max}$ = .95<br>$Genes_{cov}$ = .88<br>$Env_{cov}$ = .98<br>$Main_{cov}$ = .81 |
| **Example 2**:<br>*3 Way interaction model*<br>4 genes, 2 gene × gene interaction, 3 environments | $R^2_{val}$ / $R^2_{max}$ = .97<br>$Genes_{cov}$ = .91<br>$Env_{cov}$ = .96<br>$Main_{cov}$ = .91 | $R^2_{val}$ / $R^2_{max}$ = .97<br>$Genes_{cov}$ = .91<br>$Env_{cov}$ = .96<br>$Main_{cov}$ = .91 | $R^2_{val}$ / $R^2_{max}$ = .90<br>$Genes_{cov}$ = .85<br>$Env_{cov}$ = .96<br>$Main_{cov}$ = .89 | $R^2_{val}$ / $R^2_{max}$ = .90<br>$Genes_{cov}$ = .85<br>$Env_{cov}$ = .96<br>$Main_{cov}$ = .89 |
| Large sample (N = 5000) | | | | |
| **Example 1**:<br>*2 Way interaction model*<br>4 genes, 2 gene × gene interactions, 3 environments | $R^2_{val}$ / $R^2_{max}$ = 1<br>$Genes_{cov}$ = .93<br>$Env_{cov}$ = .98<br>$Main_{cov}$ = .90 | $R^2_{val}$ / $R^2_{max}$ = 1<br>$Genes_{cov}$ = .93<br>$Env_{cov}$ = .97<br>$Main_{cov}$ = .90 | $R^2_{val}$ / $R^2_{max}$ = .99<br>$Genes_{cov}$ = .93<br>$Env_{cov}$ = .97<br>$Main_{cov}$ = .90 | $R^2_{val}$ / $R^2_{max}$ = .99<br>$Genes_{cov}$ = .93<br>$Env_{cov}$ = .97<br>$Main_{cov}$ = .90 |
| **Example 2**:<br>*3 Way interaction model*<br>4 genes, 2 gene × gene interaction, 3 environments | $R^2_{val}$ / $R^2_{max}$ = .99<br>$Genes_{cov}$ = .92<br>$Env_{cov}$ = .98<br>$Main_{cov}$ = .87 | $R^2_{val}$ / $R^2_{max}$ = .99<br>$Genes_{cov}$ = .92<br>$Env_{cov}$ = .98<br>$Main_{cov}$ = .87 | $R^2_{val}$ / $R^2_{max}$ = .98<br>$Genes_{cov}$ = .90<br>$Env_{cov}$ = .98<br>$Main_{cov}$ = .89 | $R^2_{val}$ / $R^2_{max}$ = .98<br>$Genes_{cov}$ = .91<br>$Env_{cov}$ = .98<br>$Main_{cov}$ = .89 |

*Notes*: For example 1, we used Gaussian noise with mean 0 and standard deviation equal to 4.36 for the medium effect size ($R^2$ = .30) and equal to 6.78 for the small effect size ($R^2$ = .15).
For example 2, we used Gaussian noise with mean 0 and standard deviation equal to 12.31 for the medium effect size ($R^2$ = .30) and equal to 19.19 for the small effect size ($R^2$ = .15).

**Table 3:** *The prediction of negative emotionality (NE) at 3, 6, 18 and 36 months of age from the interaction of prenatal maternal depression and a multi-locus genetic score (longitudinal mixed model analysis with a continuous outcome, N=607).*

| Predictors | Traditional | | **Alternating Optimization** | |
|---|---|---|---|---|
| | Covariates | 2-way | Genetic | OXT |
| Number of free parameters | 6 | 9 | 10 | 11 |
| G×E | | | | |
| Intercept at 3 and 6 Months (IBQ) | .08 $^t$ | .21 | .26 $^t$ | .41* |
| Intercept at 18 and 36 Months (ECBQ) | -.15 | .03 | .07 | .23 |
| Prenatal depression | | .00 | .00 | -.01 |
| Genetic score | | -.42** | -.45** | .76*** |
| Prenatal depression × genetic score | | .02 $^t$ | .02* | .05** |
| Genetic score weights | | | | |
| DRD4 (6, 7 or 8 repeats) | | .50 (fixed) | .33 $^t$ | .19 $^t$ |
| 5-HTTLPR (S/Lg) | | .50 (fixed) | .67*** | .44*** |
| OXT (AC or AA) | | | | .37*** |
| Covariates (inside G×E) | | | | |
| Postnatal depression (At latest previous time point) | .02*** | .01*** | .01*** | .01*** |
| College education | | | | |
| Material/Social deprivation index (Quintile) | .20** | .16* | .16** | .16* |
| | -.05* | -.07** | -.07** | -.06** |
| Mother's age of birth | -.01* | -.01* | -.01* | -.02* |
| **In-sample $R^2$** | .11 | .17 | .18 | **.20** |
| **Out of sample $R^2$ (Leave-one-out)** | .08 | .13 | .13 | **.14** |
| **AIC** | 921.63 | 901.71 | 902.48 | **892.48** |
| **BIC** | 974.57 | 964.56 | 968.38 | **954.22** |

$^t p < .10$, $*p < .05$, $**p < .01$, $***p < .001$

**Table 4:** *The prediction of attention from 18 to 24 months of age from the interaction of prenatal maternal depression, maternal sensitivity and a multi-locus dopaminergic genetic score (longitudinal mixed model analysis with a continuous outcome, N=212).*

| Predictors | Traditional | | **Alternating optimization** | | |
|---|---|---|---|---|---|
| | Covariates | 3-way | Sensitivity | Genetic | G × G |
| Number of free parameters | 4 | 11 | 13 | 16 | 18 |
| G×E$_1$×E$_2$ | | | | | |
| Intercept at 18 Months | .96*** | .99*** | 1.00*** | 1.02*** | 1.06*** |
| Intercept at 24 Months | 1.38*** | 1.15 *** | 1.20*** | 1.23*** | 1.26*** |
| Prenatal depression | | .01$^t$ | .01$^t$ | .01* | .01 |
| Genetic score | | .62** | .59** | .55** | .99** |
| Maternal sensitivity score | | .35* | .59* | .42* | .26 |
| Prenatal depression × genetic score | | -.04** | -.05** | -.05*** | -.09*** |
| Prenatal depression × maternal sensitivity score | | -.04*** | -.08*** | -.07*** | -.04*** |
| Genetic score × maternal sensitivity score | | -.68** | -1.27* | -.93*** | -1.11* |
| Prenatal depression × genetic score × maternal sensitivity score | | .10*** | .24*** | .20*** | .31*** |
| Genetic score weights | | | | | |
| DRD4 (6, 7 or 8 repeats) | | .25 (fixed) | .25 (fixed) | .09 | .12* |
| DAT1 (10/10) | | .25 (fixed) | .25 (fixed) | .41*** | .26*** |
| BDNF (Val/Val) | | .25 (fixed) | .25 (fixed) | .25*** | .05*** |
| COMT (Met/Met) | | .25 (fixed) | .25 (fixed) | .25*** | .05*** |
| DRD4 × DAT1 | | | | | -.20** |
| DRD4 × COMT | | | | | .32*** |
| Maternal sensitivity score weights | | | | | |
| Looking away - percentage (Inattention) | | -1 (fixed) | -.42*** | -.45*** | -.45*** |
| Kissing - percentage (Tactile) | | | | | |
| Play without toy - frequency (Play) | | | .37*** | .34*** | .34*** |
| | | | .21*** | .21*** | .21*** |
| Covariates (inside G×E$_1$×E$_2$) | | | | | |
| Postnatal depression (12 Months) | -.01* | -.01$^t$ | -.01$^t$ | -.01$^t$ | -.01 |
| Regulation at 6 Months (IBQ) | .14* | .14** | .10* | .12* | .14** |
| **In-sample $R^2$** | .12 | .23 | .35 | .38 | **.41** |
| **Out of sample $R^2$ (Leave-one-out)** | .08 | .12 | .24 | .20 | **.25** |
| **AIC** | 242.56 | 234.97 | 213.11 | 213.39 | **208.28** |
| **BIC** | 262.07 | 274 | 256.51 | 260.05 | **255.64** |

$^t p < .10$, $*p < .05$, $**p < .01$, $***p < .001$

## Appendix A: Details of the alternating optimization algorithm

In step 1, we set the starting points; for the first time, we can simply use $\widehat{\boldsymbol{p}}^0 = \left(\frac{1}{k}, \dots, \frac{1}{k}\right)$ and $\widehat{\boldsymbol{q}}^0 = \left(\frac{1}{s}, \dots, \frac{1}{s}\right)$. Step 2.1 is trivial, one must simply fit the model from equation (2) with $\boldsymbol{p} = \widehat{\boldsymbol{p}}^i$ and $\boldsymbol{q} = \widehat{\boldsymbol{q}}^i$ using standard mixed model algorithms.

Step 2.3 is simple but requires a little bit of algebra. Assuming that:

$r_0 = \beta_0 + \beta_e \boldsymbol{e} + \boldsymbol{X_{covs}\beta_{covs}}$ ,
$r_1 = \beta_g + \beta_{eg}\boldsymbol{e}$.

We can rewrite equation (2) in the following way:

$$\boldsymbol{y} = r_0 + r_1\boldsymbol{g} + \boldsymbol{\varepsilon}.$$

Knowing that $\boldsymbol{g} = \sum_{j=1}^{k} p_j \boldsymbol{g_j}$we can reformulate the problem in terms of solving for the genetic variants:

$$(\boldsymbol{y} - r_0) = \sum_{j=1}^{k} p_j (r_1 \boldsymbol{g}_j) + \boldsymbol{\varepsilon}. \quad (5)$$

Reparametrizing the equation so that $\boldsymbol{y}' = (\boldsymbol{y} - r_0)$ and $\boldsymbol{r}_1^j = (r_1\boldsymbol{g}_j)$, we see that the model in Step 2.3 can be represented by the following equation:

$$\boldsymbol{y}' = \sum_{j=1}^{k} p_j \boldsymbol{r}_1^j + \boldsymbol{\varepsilon}.$$

where $\boldsymbol{r}_1^1, \dots, \boldsymbol{r}_1^k$ are the variables and $\boldsymbol{y}'$ is the outcome. This is easily solvable using the same algorithm used in Step 2.1.

Similarly, we solve step 2.5 assuming that:

$r_0{}' = \beta_0 + \beta_g \boldsymbol{g} + \boldsymbol{X_{covs}\beta_{covs}}$ ,
$r_1{}' = \beta_e + \beta_{eg}\boldsymbol{g}$.

We can rewrite equation (2) in the following way:

$$\boldsymbol{y} = r_0{}' + r_1{}'\boldsymbol{e} + \boldsymbol{\varepsilon}.$$

Knowing that $\boldsymbol{e} = \sum_{l=1}^{s} q_l \boldsymbol{e_l}$ we can reformulate the problem in terms of solving for the genetic variants:

$$(\boldsymbol{y} - r_0{}') = \sum_{l=1}^{s} q_l (r_1{}'\boldsymbol{e_l}) + \boldsymbol{\varepsilon}. \quad (6)$$

Reparametrizing the equation so that $\boldsymbol{y}' = (\boldsymbol{y} - r_0{}')$ and $\boldsymbol{r}_1^l{}' = (r_1{}'\boldsymbol{e}_l)$, we see that the model in Step 2.3 can be represented by the following equation:

$$\boldsymbol{y}' = \sum_{l=1}^{s} q_l \boldsymbol{r}_1^l{}' + \boldsymbol{\varepsilon}.$$

where $\boldsymbol{r}_1^1{}', \dots, \boldsymbol{r}_1^s{}'$ are the variables and $\boldsymbol{y}'$ is the outcome. Again, this can be easily solved using the same algorithm as in Step 2.1. In step 2.7, the convergence threshold $\delta$ can be chosen to be any small number. As a rudimentary guideline, for quick simulations when precision is not important, one can choose $\delta = .01$ and for longer and more precise simulations, one can choose $\delta = .0001$.

## Appendix B: Notes and recommendations

### Direction and local convergence

Alternating optimization only guarantees convergence to a local optimum, therefore the algorithm may provide a suboptimal solution. Experimentally, we noticed that suboptimal convergence happened mostly when, for more than one genetic variant or environmental factor, the direction of the effect was the opposite of the starting point. We, therefore, recommend adding genetic variants or environments one by one rather than adding multiple variables simultaneously. This way, if an additional genetic variant (assumed to be binary) is found to have a negative weight, it can either be recoded as $(1 - \boldsymbol{g}_j)$ or the starting point direction of effect can be reversed to prevent the possibility of some weights going in the wrong direction. Similarly, if an additional environmental exposure (assumed to be continuous) is found to have a negative weight, it can either be recoded as $(-\boldsymbol{e}_l)$ or the starting point direction of effect can be reversed. Adding variables one by one is also preferable to prevent multicollinearity and overfitting.

### Overfitting and variable selection

Including every statistically significant genetic variant or environmental factor into the genetic/environmental score would rapidly lead to an excess of variables and overfitting. Instead, we recommend only adding variables that decrease the out-of-sample cross-validated error (or equivalently increase the out-of-sample cross-validated $R^2$) or optimize a log-likelihood based criterion of model fit, like the AIC or BIC.

One must be careful though with cross-validation or a log-likelihood based criterion when using the alternating optimization approach. One must make sure to create the cross-validation folds before running the alternating optimization algorithm, otherwise, the weights of the genetic or environmental scores would be incorrectly assumed to be the same in every fold. For log-likelihood based criterion, one cannot rely on the output given by the software as it won't account for the variables used in the genetic/environmental scores. Instead, one must take the log-likelihood and calculate manually the desired criterion based on the true number of parameters inside the model, thus considering the variables inside the genetic and environmental scores.

One cannot easily use traditional regularization techniques like lasso (Tibshirani, 1996) or elastic net (Zou and Hastie, 2005) with the alternating optimization approach due to the non-linear nature of the objective function. Applying regularization to the parts of the models that assume other parameters to be known would also be problematic because the solution would be conditional on the other parameters being known and convergence of the alternating optimization sequence wouldn't be guaranteed anymore. To do variable selection in an automated way, we instead recommend using a stepwise approach (forward, backward or bidirectional) based on log-likelihood criteria or cross-validation.

## Outliers

Outlier detection can be challenging with the alternating optimization approach as one can never see the full model but only parts of it while holding other parameters constant. Initially, we considered an observation to be an outlier if the studentized residual was greater than 2.8 (probability of .005) or greater than 2 (probability of .05) with a combined leverage larger than 2p/n (Hoaglin and Welsch, 1978). This strategy worked well to identify outliers in longitudinal mixed models.

With the alternating optimization approach we have three models to estimate; Step 2.1 which estimate $\boldsymbol{\beta}$ while holding $\boldsymbol{p}$ and $\boldsymbol{q}$ constant, Step 2.3 consisting which estimate $\boldsymbol{p}$ while holding $(\boldsymbol{\beta}, \boldsymbol{q})$ constant and Step 2.5 which estimate $\boldsymbol{q}$ while holding $(\boldsymbol{\beta}, \boldsymbol{p})$ constant. We initially thought that we could apply our standard strategy for outlier detection in all three models and remove all observations that were detected as outliers in any of the three models. However, this strategy worked very poorly, as we were unable to detect all outliers, leading to the removal of too many observations (each of the three models can have different outliers). For example, in one model there was a single individual that reduced the out-of-sample effect size from .2 to .13. Although this individual was clearly an outlier, it was not detected by the algorithm. Based on these results, we starting using a different approach to categorize outliers.

What we recommend instead is to classify outliers by looking at the standardized cross-validated leave-one-out cross-validated (LOOCV) residuals. A threshold > 2.8 (p = .005) can be chosen for a conservative classification and > 2.5 (p=.01) can be used for a more optimistic classification. For GLMs, the Pearson residuals or the deviance residuals can be used instead of standard residuals.

## Equivalent parameterizations

It is important to note that with one genetic score and one environmental score, there are 4 possible parameterizations to represent the same model:

1) $\boldsymbol{y} = \beta_1 + \beta_G \boldsymbol{g} + \beta_G \boldsymbol{e} + \beta_{GE} \boldsymbol{ge} + \boldsymbol{\varepsilon}$ with $\boldsymbol{g} = \sum_{j=1}^{k} p_j \boldsymbol{g_j}$ and $\boldsymbol{e} = \sum_{l=1}^{s} q_l \boldsymbol{e_l}$ ,
2) $\boldsymbol{y} = \beta_1 - \beta_G \boldsymbol{g} + \beta_G \boldsymbol{e} - \beta_{GE} \boldsymbol{ge} + \boldsymbol{\varepsilon}$ with $\boldsymbol{g} = -\sum_{j=1}^{k} p_j \boldsymbol{g_j}$ and $\boldsymbol{e} = \sum_{l=1}^{s} q_l \boldsymbol{e_l}$ ,
3) $\boldsymbol{y} = \beta_1 + \beta_G \boldsymbol{g} - \beta_G \boldsymbol{e} - \beta_{GE} \boldsymbol{ge} + \boldsymbol{\varepsilon}$ with $\boldsymbol{g} = \sum_{j=1}^{k} p_j \boldsymbol{g_j}$ and $\boldsymbol{e} = -\sum_{l=1}^{s} q_l \boldsymbol{e_l}$ ,
4) $\boldsymbol{y} = \beta_1 - \beta_G \boldsymbol{g} - \beta_G \boldsymbol{e} + \beta_{GE} \boldsymbol{ge} + \boldsymbol{\varepsilon}$ with $\boldsymbol{g} = -\sum_{j=1}^{k} p_j \boldsymbol{g_j}$ and $\boldsymbol{e} = -\sum_{l=1}^{s} q_l \boldsymbol{e_l}$ .

This complicates the inspection of the fitted model parameters because if the true model has parameterization 1 but the fitted model has parameterization 2, 3 or 4, it will appear as if the parameters are incorrect even though they are actually correct. This does not affect the predictions because the parameterizations are equivalent. However, this might affect the coverage (Dodge, 2006), e.g. the percentage of times that a parameter's confidence interval contains its true value because some of the coefficients might be inverted. Not knowing the true parameters of the model, it is impossible to know which parameterization is the original one. Even knowing the original parameterization, for models with a lot of parameters like $G \times E_1 \times E_2$, it can still be hard to know which of the parameterizations best represents the fitted model. Therefore, we recommend considering the "true" coverage of a parameter to be the one obtained from the parameterization for which the average coverage of all parameters was the highest. Note that if there were $M$ genetic scores and $J$ environmental scores, there would be $2^{M+J}$ equivalent parameterizations.

## Appendix C: Details of the model's variations

### Model with three-way interactions G×E₁×E₂

So far we have only looked at the basic G×E model. However, often the genetic effects are thought to depend on multiple unrelated environmental exposures. In the example of two environmental factors, this model is best represented as a three-way $G \times E_1 \times E_2$ model:

$$\boldsymbol{y} = \beta_0 + \beta_1 \boldsymbol{e}_1 + \beta_2 \boldsymbol{e}_2 + \beta_3 \boldsymbol{g} + \beta_{12} \boldsymbol{e}_1 \boldsymbol{e}_2 + \beta_{13} \boldsymbol{e}_1 \boldsymbol{g} \\ + \beta_{23} \boldsymbol{e}_2 \boldsymbol{g} + \beta_{123} \boldsymbol{e}_1 \boldsymbol{e}_2 \boldsymbol{g} + \boldsymbol{X}_{covs} \boldsymbol{\beta}_{covs} + \boldsymbol{\varepsilon}\,,$$

where $\boldsymbol{e}_1$ and $\boldsymbol{e}_2$ are the environmental scores and $\boldsymbol{g}$ is the genetic score.

To construct the three-way model, one would have to make the following modifications to the algorithm:

- Add two extra steps between Step 2.6 and Step 2.7 for the estimation of $\boldsymbol{e}_2$ and its division by the sum of the absolute values of the weights.
- change $r_0$ and $r_1$ in equation (5) of Step 2.3 to include the terms with $\boldsymbol{e}_2$
- change $r_0{}'$ and $r_1{}'$ in equation (6) of Step 2.5 to include the terms with $\boldsymbol{e}_2$.

### Mixed model

Assuming a two-way interaction between the genetic score $\boldsymbol{g}$ and the environmental score $\boldsymbol{e}$, the mixed model can be defined as:

$$\boldsymbol{y} = \beta_0 + \beta_e \boldsymbol{e} + \beta_g \boldsymbol{g} + \beta_{eg} \boldsymbol{eg} + \boldsymbol{X}_{covs} \boldsymbol{\beta}_{covs} + \mathbf{Z}\boldsymbol{\gamma} + \boldsymbol{\varepsilon}\,, \tag{2}$$

where $\boldsymbol{y}$ is a vector representing the $n$ observed outcomes, $\beta_0, \beta_e, \beta_g, \beta_{eg}$ are scalars of the unknown fixed-effect parameters for the G×E, $\boldsymbol{X}_{covs}$ is a design matrix for additional fixed-effects, $\boldsymbol{\beta}_{covs}$ a vector of unknown additional fixed-effect parameters, $\mathbf{Z}$ a design matrix for the random-effects, $\boldsymbol{\gamma}$ a vector of unknown random-effect parameters, $\boldsymbol{\varepsilon}$ is a Gaussian noise and $E\begin{bmatrix}\boldsymbol{\gamma}\\ \boldsymbol{\varepsilon}\end{bmatrix} = \begin{bmatrix}0\\0\end{bmatrix}, Var\begin{bmatrix}\boldsymbol{\gamma}\\ \boldsymbol{\varepsilon}\end{bmatrix} = \begin{bmatrix}\mathbf{D} & 0\\ 0 & \boldsymbol{R}\end{bmatrix}$, where $\mathbf{D}$ is an unknown covariance matrix for the random effects and $\boldsymbol{R}$ is an unknown covariance matrix for the residuals (with repeated measures, the residuals are not independent). Note that setting $\boldsymbol{Z} = 0$ and $\boldsymbol{R} = \sigma^2 \mathrm{I}_n$ leads to the usual multiple linear model. The algorithm can be rewritten in the following way:

| **Algorithm 2**: Alternating optimization for estimating the parameters of a two-way G×E mixed model. |
|---|
| 1. Set $\widehat{\boldsymbol{p}}^0$ and $\widehat{\boldsymbol{q}}^0$ to reasonable starting points<br>2. Until convergence, for $i = 1$ to $max_iterations$<br>2.1. Estimate $(\boldsymbol{\beta}, \mathbf{D}, \boldsymbol{R})$ assuming $\boldsymbol{p} = \widehat{\boldsymbol{p}}^{i-1}$ and $\boldsymbol{q} = \widehat{\boldsymbol{q}}^{i-1}$ |

2.2. Let $(\hat{\boldsymbol{\beta}}^i, \hat{\boldsymbol{D}}^i, \hat{\boldsymbol{R}}^i) = (\hat{\boldsymbol{\beta}}, \hat{\mathbf{D}}, \hat{\boldsymbol{R}})$
2.3. Estimate $\boldsymbol{p}$ assuming$(\boldsymbol{\beta}, \mathbf{D}, \boldsymbol{R}) = (\hat{\boldsymbol{\beta}}^i, \hat{\boldsymbol{D}}^i, \hat{\boldsymbol{R}}^i)$ and $\boldsymbol{q} = \hat{\boldsymbol{q}}^{i-1}$
2.4. Let $\hat{\boldsymbol{p}}^i = \frac{\hat{\boldsymbol{p}}}{\|\hat{\boldsymbol{p}}\|_1}$
2.5. Estimate $\boldsymbol{q}$ assuming$(\boldsymbol{\beta}, \mathbf{D}, \boldsymbol{R}) = (\hat{\boldsymbol{\beta}}^i, \hat{\boldsymbol{D}}^i, \hat{\boldsymbol{R}}^i)$ and $\boldsymbol{p} = \hat{\boldsymbol{p}}^i$
2.6. Let $\hat{\boldsymbol{q}}^i = \frac{\hat{\boldsymbol{q}}}{\|\hat{\boldsymbol{q}}\|_1}$
2.7. If $\|\hat{\boldsymbol{p}}^i - \hat{\boldsymbol{p}}^{i-1}\| < \delta$ and $\|\hat{\boldsymbol{q}}^i - \hat{\boldsymbol{q}}^{i-1}\| < \delta$ then convergence is attained (break loop)
3. Return $(\hat{\boldsymbol{\beta}}^i, \hat{\boldsymbol{D}}^i, \hat{\boldsymbol{R}}^i, \hat{\boldsymbol{p}}^i, \hat{\boldsymbol{q}}^i)$

In both Step 2.3 and Step 2.5, the covariance matrices $\mathbf{D}$ and $\boldsymbol{R}$ must be held constant; this can be done in SAS with PROC MIXED and PROC GLIMMIX. No currently existing R package seems to possess this capability.

**Non-identity GLM link function**

In a generalized linear mixed models (GLMM) set-up, equation (2) would now be represented as:

$$\boldsymbol{E}(\boldsymbol{y}|\boldsymbol{\gamma}) = g^{-1}(\boldsymbol{X\beta} + \mathbf{Z}\boldsymbol{\gamma}) \tag{7}$$

$$\boldsymbol{Var}(\boldsymbol{\gamma}) = \boldsymbol{D}$$

$$\boldsymbol{Var}(\boldsymbol{y}|\boldsymbol{\gamma}) = \boldsymbol{A}^{1/2}\boldsymbol{R}\boldsymbol{A}^{1/2}$$

where $g$ is the link function and $\boldsymbol{A}$ is a diagonal matrix containing the variance function, the variance of the outcome as a function of the mean and the scale parameter $\phi$, a constant to be estimated ($\phi = \sigma^2$ for normal distribution) or known ($\phi = 1$ for binomial and Poisson distribution). The mean and variance of the outcome depends on the assumed distribution. Assuming that $\boldsymbol{y}$ has a normal distribution and using an identity link function lead to the standard linear mixed model as in equation (2). Assuming that $\boldsymbol{y}$ has a binomial distribution and using a logit link function leads to the logistic mixed model. Setting $\boldsymbol{Z} = 0$ and $\boldsymbol{R} = \mathrm{I}_n$ leads to the standard GLM.

Step 2.1 is still trivial and equivalent to solving equation (7) while holding $\boldsymbol{p}$ and $\boldsymbol{q}$ constant. Steps 2.3 and 2.5, on the other hand, become a bit more complicated. In step 2.3, equation (5) in a GLM setting is now:

$$\boldsymbol{E}(\boldsymbol{y}|\boldsymbol{\gamma}) = g^{-1}(r_0 + \textstyle\sum_{j=1}^{k} p_j(r_1\boldsymbol{g}_j) + \mathbf{Z}\boldsymbol{\gamma}). \tag{8}$$

To fit this model, one must include $r_0$ as an offset, a variable for which the β estimate is set to 1. Options for setting up offsets are available in PROC LOGISTIC and PROC GLIMMIX in SAS. The same idea can be applied in Step 2.5 to fit the model in equation (6) when we have a GLM link function. Without access to an offset option, one could theoretically apply a transformation to the outcome and fit the model with the transformed outcome, just like we did with the linear mixed model where we transformed the outcome as $\boldsymbol{y}' = (\boldsymbol{y} - r_0)$ in equation (5) and $\boldsymbol{y}' = (\boldsymbol{y} - r_0{}')$ in equation (6).. The transformed outcome could end up being a complex number with a non-zero imaginary part though, thus making it impossible to set up; this is the case with the logistic mixed model.